\documentstyle[prd,aps,floats]{revtex}
\begin{document}
\draft

\input epsf \renewcommand{\topfraction}{0.8} 
\twocolumn[\hsize\textwidth\columnwidth\hsize\csname 
@twocolumnfalse\endcsname

\title{ A Note on Brane Inflation} 
\author{Anupam Mazumdar$^a$ and Jing Wang$^b$}
\address{$^a$Astrophysics Group, Blackett Laboratory, Imperial College London,
 SW7 2BZ, ~~~U.~K.\\
$^b$Theory Division, Fermi National Accelerator Laboratory\\
P.O.Box 500, Batavia, IL 60510, USA\\}
\date{\today} 
\maketitle
\begin{abstract}
We demonstrate that there exists an inflationary solution on the positive tension 
brane in the Randall-Sundrum scenario. Inflation is driven by a slow-rolling scalar field 
on the brane and is achieved within the perturbative limit of the radion field. 
We find that inflation on the positive tension brane results in a slight increase in the 
separation between the two branes. 
However, we show that the slow-roll inflation is not possible on the 
negative tension brane.  
       
\end{abstract}

\pacs{PACS numbers: 98.80.Cq \hspace*{1.3cm} FERMILAB-Pub-00/076-T, Imperial preprint Imperial-AST 
00/4-2, hep-ph/yymmmnn}

\vskip2pc]

\section{Introduction}
Recently, an alternative mechanism of solving the hierarchy problem 
has been proposed \cite{nima}. The novel proposal exploits the fact 
that the fundamental Planck scale $M$ is a TeV scale but in $4+d$
dimensions,
where $d$ is the number of extra dimensions which are to be compactified. 
The Planck scale in the observable world is generated because of the presence
of large extra dimensions: $M_{\rm Pl}^2 = M^{2+d}V_d$. Here
$V_d$ is the volume of the extra compactified dimensions. The proposal
also demands that the standard model fields are bound to live in the 
observable world and gravity is the only force which mediates through 
the bulk and the observable world. The introduction of very large 
size of the compactified dimensions  
brings a completely new set of exciting problems ranging from
phenomenology of accelerator physics \cite{lykken} to cosmology \cite{numero}.   

However, Randall and Sundrum \cite{rs} introduced another twist in the physics beyond $4$ dimensions to 
account for the hierarchy between the Planck scale and the electroweak scale.
They demonstrate that in a background of 
special non-factorizable geometry an exponential warp factor appears for the Poincar\'e invariant 3+1 dimensions.
The model consists of
two $3$ branes situated at the fixed positions along the 5th dimension compactified on a $S^{1}/Z_2$ 
orbifold symmetry. The space-time in the bulk is 5 dimensional anti de-Sitter space (AdS).  The 
five dimensional Einstein's equations permit a solution which preserves 4 dimensional Poincar\'e 
invariance on the brane, with the metric taking the form 
\begin{eqnarray}
ds^2 = e^{-2k|z|}g_{\mu \nu}dx^{\mu}dx^{\nu}+dz^2 \,.
\end{eqnarray}
Here $z$ is the extra spatial dimension and $\mu, \nu =0,..,3$. The constant $k$ is 
determined by the bulk cosmological constant $-\Lambda=3M_5^3k^2$ where
$M_5$ is the five dimensional Planck scale (see the 5 dimensional action 
defined later). The 4 dimensional Poincar\'e invariance 
also requires a fine-tuning of the brane tensions. Namely, the positive tension $\sigma$ 
of the brane at $z=0$ is related to $\Lambda$ and $M_5$, while
the brane at $z=z_c$ has equal and opposite tension $-\sigma$.  Gravity is localized due 
to the exponentially decaying warp factor in the 5th dimension. The hierarchy between 
the Planck scale and the electroweak scale is explained by the 
suppression factor $e^{-kz_c}$ on the negative tension brane, 
where the standard model (SM) particles are assumed to be.   

The excitations around the background metric includes a massless 4 dimensional graviton 
zero modes and a set of massive Kaluza-Klein modes which have masses proportional to $1/z_c$. 
In addition, there is a massless four-dimensional scalar associated with the relative motions 
of the two branes, i.e., the radion field \cite{charmousis}. The radion field would become 
massive once the separation  of the two branes are stabilized due to certain mechanism \cite{gw}. 
It is interesting to note that the reduced action of the massless graviton zero mode and the radion field   
in the effective $4$ dimensions on either of the branes, is not exactly the same as 
general relativity (GR). Instead, the actions on the positive and the negative branes mimic that of
a deviant theory of gravity, popularly known as scalar tensor theory, with the radion field taking 
the role of a Brans-Dicke scalar with non-trivial Brans-Dicke parameter \cite{chiba}. 
In this theory, Einstein's  gravity is recovered in a limiting case. 

The purpose of this paper is to investigate the possibility of inflationary scenarios on the branes. 
We assume that the inflation is induced by a scalar field confined on the brane which has 
a slow-roll potential. In our discussion, we also assume that the inflation on the brane 
takes place before the stabilization of the extra dimensions \cite{adkm:inflation}, hence 
during the inflationary era, the radion field is massless and has no additional potential. 
We find that it is possible to have an inflationary era induced by a slow-rolling inflaton 
on the positive tension brane. In particular, during this era the separation between the 
two branes increases. Namely, the bulk gravity reacts to inflation on the Planck
 brane and pushes the other brane away and enlarges the size of the AdS space. We find
 that the change in the size of the extra dimension can be sufficiently small such that 
the perturbative description of our frame work is still valid towards the end of inflation.  
However, on the negative tension brane, we fail to find a consistent inflationary solution 
within our assumptions of massless radion and slow-rolling inflaton. 

In Section II, we write down the effective actions on the branes. We present the equations 
of motions in the next section and the inflationary solutions on both the positive and 
the negative tension branes. In Section IV, we discuss our results.

\section{Effective actions}

The full action consists of $5$ dimensional Einstein's gravity
with a negative cosmological constant in the bulk. The two branes are located
on the orbifold $S^{1}/Z_2$ along the extra dimension, with the
positive tension brane at $z=0$ and the negative tension brane at $z=z_c$. 
\begin{eqnarray}
\label{action}
S = 2\int d^4 x \int_0^{z_c}dz \sqrt{-g_5}\left[\frac{M_5^3}{2}R_5
-2\Lambda\right]\,
\nonumber \\
-\sigma_{+}\int d^4x \sqrt{-g_{+}}-\sigma _{-}\int d^4x\sqrt{-g_{-}}\,,
\end{eqnarray}
where $M$ is the $5$ dimensional Planck mass, $\sigma_{\pm}$ and $g_{\pm}$ are the
brane tensions and the corresponding induced metric on the respective branes. The effective 4 
dimensional Planck scale: $M_{Pl}=M_5^3(1-e^{-2kz_c})/(4k)$ \cite{rs}.
The linearized gravity around its background including the massless degree of freedom, i.e., 
the massless 4 dimensional graviton and radion field, can be parametrized by the 
following metric solution \cite{charmousis}, 
\begin{eqnarray}
ds^2 &=& e^{-2kh(x,z)}\bar g_{\mu \nu}dx^{\mu}dx^{\nu} +h^2_{,z}dz^2 \,, \nonumber \\
h(x,z)&=& z+f(x)e^{2kz}\,.
\end{eqnarray} 
where $0 < z <z_c$, and $h_{,z}$ denotes the derivative with respect to $z$. 
The radion field $f(x)$, which is a function of the brane coordinates only, 
was also introduced in \cite{csaki}. For the classical solution, 
the induced metric on the positive tension brane at $z=0$ takes the form,
\begin{eqnarray}
\label{g+}
g_{+ \mu \nu}= e^{-2kf}\bar g_{\mu \nu}\,,
\end{eqnarray}
and on the negative tension brane at $z=z_c$:
\begin{eqnarray}
\label{g-}
g_{- \mu \nu}=\alpha^{-1} e^{-2\alpha kf}\bar g_{\mu \nu}\,,
\end{eqnarray}
where $\alpha \equiv e^{2kz_c}$ is a constant. With this metric ansatz one 
can integrate out the fifth dimension to achieve the effective four 
dimensional actions on the two branes \cite{chiba}.
\begin{eqnarray}
\label{S}
S_{\pm}=\int d^4x \sqrt{-g_{\pm}}\frac{1}{16\pi}\left[\Phi^{\pm}R-
\frac{\omega_{\pm}(\Phi^{\pm})}{\Phi^{\pm}}\nabla (\Phi^{\pm})^2\right]\,,
\end{eqnarray}
where, $\pm$ denotes the positive and negative tension branes respectively. The 
fields $\Phi^{\pm}$ and the parameter $\omega_{\pm}$ (dimensionful) are 
defined as \cite{chiba}, 
\begin{eqnarray}
\label{Phi}
\Phi^{\pm}=\frac{2}{kG_5}e^{\mp k[(\alpha -1)f+z_c]}\sinh (k[(\alpha-1)f +z_c])
\,,
\end{eqnarray}
\begin{eqnarray}
\label{omega}
\omega_{\pm}(f)=\pm 3e^{\pm k[(\alpha -1)f+z_c]}\sinh (k[(\alpha -1)f+z_c])\, \nonumber \\
\left(
1\mp \frac{e^{\pm k[(\alpha -1)f]}\sinh [k(\alpha -1)f]}{\sinh^2 (kz_c)}\right)\,,
\end{eqnarray}
where $G_5$ is defined as $G_5\equiv (8\pi M_5^3)^{-1}$. The effective
action (\ref{S}) is the 
correct description of the dynamics 
of the massless fields as long as the perturbative limit 
$e^{2k\alpha f} \ll \alpha$ is satisfied \footnote{Here we want to emphasis that in the strict limit $e^{2k\alpha f} \ll \alpha$, the effective action on the brane (\ref{S}) is equivalent to the original action in five dimensions. It is possible that effective brane action deduced from dimensional reduction could differ in higher orders from that of the original 5-dimensional action \cite{TS}.}. It is pointed 
out in \cite{chiba} that this action is similar to that
of the scalar tensor theories \cite{barrow0}, with $\Phi^{+}$ and 
$\Phi^{-}$ are the scalar Brans-Dicke fields,
whose dynamical evolution determines the strength of gravity, namely, 
the Newton's constant $G_{N} \equiv 1/\Phi_{\pm}$. 
The strength of the coupling between gravity and the scalar field $\Phi^{\pm}$ is determined by
$\omega_{\pm}$. When $\omega_{\pm} \rightarrow \infty$, the scalar tensor theory approaches 
Einstein's general theory of relativity provided $\omega^{\prime}_{\pm}/\omega_{\pm}^{ 3} \rightarrow 0$ 
\cite{nordt}. The present observations such as 
light bending, perihelion precession and radar echo delay phenomena in the solar system 
 suggests $\omega_{\pm} > 3000$  
\cite{will}. It is obvious that in the perturbative limit $\omega_{+}$ on the positive 
tension brane $\omega_{+} \sim 3\alpha e^{2k\alpha f}/2$ can easily satisfy the 
experimental constraint, while on the negative tension brane $\omega_{-} \sim -3/2$ 
does not satisfy this bound \cite{chiba}. However, we take the view that the situation 
could be altered if additional massive degrees of freedom in the bulk 
associated with stabilization are included. 
In the following section we study the dynamics of both branes in
the presence 
of a scalar field inflaton with a slow-roll potential on the brane. We are
mainly interested in 
the inflationary aspects of the solutions. In the mean time, it is 
interesting 
to see if brane inflation could lead to a more physical $\omega_{-}$.


\section{Inflationary Solutions} 

To induce inflation on the brane, we add the action of a scalar field on the brane. 
We conjecture that adding an additional scalar field on both the branes will not 
affect the background classical solution of the gravity and the dynamics can be safely 
described with the usual perturbative theory \cite{charmousis}. At the end we shall see
that inflation on the positive tension brane indeed satisfies our assumption, 
while inconsistency appears on the negative tension brane. The action of a 
scalar field $\chi$ on the brane takes the form, 
\begin{equation}
\label{Smatter}
S_{m,\pm} = \int d^4x \sqrt{-g_{\pm}} [ g_{\pm \mu \nu} \partial^{\mu} 
\chi \partial^{\nu} \chi - V(\chi) ],  
\end{equation}
where $g_{\pm \mu \nu}$ are the induced metric on the branes as defined previously. 

To discuss inflation on the brane, we choose $\bar{g}_{\mu \nu}$ to be the Friedmann 
Robertson Walker (FRW) metric, $\bar{g}_{\mu \nu}=diag\{-1,~ A(t),~ A(t),~ A(t)\}$, where $A(t)$ is
the scale factor. 
In order to derive the equations of motions for the fields, one can 
perform a coordinate transformation on the brane, $d\tau = \sqrt{|g_{\pm 00}|} dt$,
 such that the metric on the branes, Eqs.(\ref{g+}-- \ref{g-}) can be brought to the 
ordinary FRW metric by defining the scale factors 
$a_{+}(\tau) \equiv e^{-kf(t)}A(t)$ and $a_{-}(\tau) \equiv \alpha^{-1/2} e^{-\alpha kf(t)}A(t)$.
We have explicitly assumed that the radion field $f$ has solely the time dependence.
Since we are interested in studying the inflationary solution, we can appropriately 
assume that the late time dynamics of $f$ does not depend on the spatial coordinates of the branes. 
Under these assumptions the equations for the scalar fields \cite{barrow} are   
\begin{eqnarray}
\label{eom1}
H^2 + H \frac{\dot{\Phi}^{\pm}}{\Phi^{\pm}} - 
\frac{\omega}{6} \frac{\dot{\Phi}^{\pm 2}}{ \Phi^{\pm 2}} \, \nonumber  \\
&=& \frac{8 \pi}{3\Phi^{\pm}} (\frac{1}{2} \dot{\chi}^2 + V(\chi)) \,, \\
\label{eom2}
\ddot{\Phi}^{\pm} + \dot{\Phi}^{\pm} 
\left[ 3H+ \frac{\dot{\omega}}{2\omega+3} \right]\, \nonumber \\
&=& 8 \pi \frac{4V(\chi) - \dot{\chi}^2}{2\omega+3} \,, \\
\label{eom3}
\ddot{\chi} + 3H \dot{\chi} + V'(\chi) &= &0 \, ,
\end{eqnarray}
where the Hubble parameter $H \equiv \dot{a}_{\pm}/a_{\pm}$, the overdot 
denotes $d/d\tau$ and the prime denotes $d/d\chi$. An inflationary epoch, in which the scale factors 
$a_{\pm}$ are accelerating, requires the scalar field $\chi$ to evolve
slowly compared to the expansion of the Universe. Thus, the following 
conditions of slow-rolling are required:
\begin{equation}
\label{slowroll1}
\ddot{\chi} \ll H \dot{\chi} \, , 
\end{equation}
\begin{equation}
\label{slowroll2}
\frac{1}{2}\dot{\chi}^2 \ll V(\chi) \, , 
\end{equation}
\begin{equation}
\label{slowroll3}
\ddot{\Phi}^{\pm} \ll H \dot{\Phi}^{\pm} \ll H^2  \Phi^{\pm}\, . 
\end{equation}
Under the slow-roll conditions, the Eqs.(\ref{eom1}--\ref{eom3}) 
can be simplified. In the following subsections, we discuss the possible solutions 
on the positive and the negative tension branes respectively. For
simplicity, we assume  
that $V(\chi)=V_0$ during the epoch when the slow-roll conditions 
Eqs.(\ref{slowroll1}--\ref{slowroll3}) are satisfied. There are other forms of 
$V(\chi)$ that one can take to satisfy the slow-roll conditions \cite{barrow0}, 
which we will not consider in this paper.

\subsection{On the positive tension brane} 

On the positive tension brane, when $e^{2k\alpha f} \ll \alpha$, $\Phi^{+} 
\approx G_{P}(1-e^{(-2k\alpha f)}/\alpha)$, 
and the Brans-Dicke parameter
$\omega_{+}$ can be written as a function of $\Phi^{+}$, 
\begin{eqnarray}
\omega_{+}(\Phi^{+}) \approx  
\frac{3}{2}\frac{\Phi^{+}}{G_{\rm P} - \Phi^{+}} \,,
\end{eqnarray} 
where $G_{P} \equiv 1/(kG_5)$. In this limit, 
the value of $\Phi^{+}$ approaches $G_{P}$.
And $\omega_{+} \sim \frac{3}{2} \alpha \gg 1$. 
Following the slow-roll approximations Eqs.(\ref{slowroll1}--\ref{slowroll3}),
Eqs.(\ref{eom1}--\ref{eom2})  can be expressed as:   
\begin{eqnarray}
\label{peom1}
3H^2 (G_{\rm P}-\Phi^{+}) \simeq 8\pi V_0 (\Phi^{+})^{-1}
 (G_{\rm P}-\Phi^{+}) \, \nonumber \\
+ \frac{3}{4}(\dot{\Phi}^{+})^2 (\Phi^{+})^{-1} \, , \\
\label{peom2}
3H \dot{\Phi}^{+}(G_{\rm P}-\Phi^{+}) + \frac{1}{2}\dot{\Phi}^{+2} 
(\Phi^{+})^{-1} G_{\rm P} \, \nonumber  \\
\simeq \frac{32}{3} \pi V_0 \frac{(G_{\rm P}-\Phi^{+})^2}{\Phi^{+}} 
\,. 
\end{eqnarray} 
By manipulating Eqs.(\ref{peom1}--\ref{peom2}) we get:
\begin{equation}
\label{peom3}
\frac{(\dot{\Phi}^{+})^2}{(\Phi^{+})} \frac{G_{\rm P}}{G_{\rm P}-\Phi^{+}} \\
\simeq 6H \left[ \frac{4}{3}H(G_{\rm P}-\Phi^{+}) - 
\dot{\Phi}^{+} \right] \, ,
\end{equation}
where the slow-roll condition Eq.(\ref{slowroll3}) has been used to neglect 
the term $\frac{3}{4}(\dot{\Phi}^{+})^2 (\Phi^{+})^{-1}$. The above 
equation can be solved approximately by: 
\begin{equation}
\label{psolv1}
\frac{\dot{\Phi}^{+}}{G_{\rm P}-\Phi^{+}} \simeq \gamma H \,,
\end{equation} 
where $\gamma = \sqrt{17}-3$, we have used Eq.(\ref{slowroll3}) and the
approximation that 
in the limit $e^{2\alpha kf} \ll \alpha$, $\Phi^{+} \sim G_{\rm P}$. With Eq.(\ref{psolv1}) and the 
slow-roll condition, Eq.(\ref{peom1}) can be reduced to 
\begin{equation}
\label{psolv2}
H \simeq \sqrt{\frac{8 \pi V_0}{3 \Phi^{+}}}. 
\end{equation}  
One can then solve for $\Phi^{+}$ from Eq.(\ref{psolv1}),  
\begin{equation}
\label{psolv3}
\begin{array}{c}
(\tau-\tau_0)\sqrt{\frac{2\pi \gamma^2}{3}\pi V_0}= ~~~~~~~~~~~~~~~~~~~~ \\
~~~~~~~~~~~~~~~~~~~\left[-\sqrt{\Phi^{+}}+\frac{\sqrt{G_{\rm P}}}{2}
\ln \frac{\sqrt{G_{\rm P}}+\sqrt{\Phi^{+}}}{\sqrt{G_{\rm P}}-\sqrt{\Phi^{+}}}
\right]^{\Phi^{+}}_{\Phi_{0}^{+}}\,,
\end{array}
\end{equation}
where $\Phi_{0}^{+}$ is the initial value of $\Phi^{+}$. We take as an initial condition 
that at the beginning of the inflation $f(\tau=\tau_0)=0$, hence $\Phi_{0}^{+}= G_{P}(1-1/\alpha)$. Solving 
Eq.(\ref{psolv1}) for the scale factor, we get:
\begin{equation}
\label{psolv4}
\frac{a(\tau)}{a_0} \simeq \left[\frac{G_{\rm P}-\Phi_{0}}{G_{P}-\Phi (t)}\right]^{1/\gamma}\,.
\end{equation}
Substituting Eq.(\ref{psolv4}) in Eq.(\ref{psolv3}) and assuming $a(\tau)/a_0 \gg 1$ at $\tau \gg \tau_0$, 
we get the final expression
\begin{equation}
\label{psolv5}
\frac{a(\tau)}{a_0} \simeq \exp\left(\sqrt{\frac{8\pi V_0}
{3G_{\rm P}}}(\tau-\tau_0)\right)\,.
\end{equation}
The above equation confirms the exponential growth in the scale factor during the inflationary era.
The evolution of the Brans-Dicke field can be obtained by substituting Eq.(\ref{psolv5})
into Eq.(\ref{psolv3}).
\begin{equation}
\label{psolv6}
\frac{\Phi^{+}(\tau)}{G_{P}} \approx  1- \frac{1}{\alpha} \exp \left(-2\left[\sqrt{\frac{2\pi \gamma^2 V_0}
{3G_{P}}(\tau-\tau_0)}\right]\right)\,, 
\end{equation}
where we have used the initial condition $f(\tau=\tau_0)=0$ and  
$\Phi_{0}^{+} = G_{P}(1-1/\alpha)$. 
Eq.(\ref{psolv6}) shows that the effective Brans-Dicke field grows with time 
during the inflationary era and approaches its asymptotic value, $(kG_5)^{-1}$ at $\tau \rightarrow \infty$.
We can also 
estimate the final value of $f(\tau_f)$ in terms of the number of 
e-foldings $N$ of the inflation, by using the expression for $\Phi_{+}$. 
From Eq.(\ref{psolv5}) and 
(\ref{psolv6}), we get:
\begin{equation}
\label{psolv7}
f(\tau_f)=\frac{\gamma N}{2k \alpha }\,,
\end{equation}
where,
\begin{eqnarray}
N\equiv \ln\left( \frac{a(\tau_f)}{a(\tau_0)}\right) \,.
\end{eqnarray}
If we assume that $N \sim 60$ e-foldings (to solve the flatness problem in the usual hot big bang Universe), 
it is not hard to realize that for sufficiently large 
$\alpha$, the condition $e^{2k\alpha f(\tau_f)} \sim e^{\gamma N} \ll \alpha$ could be satisfied such
that during the complete 
process of inflation, the perturbative description of radion field remains valid. However, 
note that the size of $\alpha$ that is required here is larger than the required size of $\alpha$ 
to explain the hierarchy problem, i.e., $e^{kz_c} \sim 10^{15}$ is needed to produce 
TeV scale masses from the fundamental scale of $M_{Pl}\sim 10^{19}$ GeV \cite{rs}. 
Since, we have assumed that the stabilization of the brane separation and the inflation 
are two independent processes taking place at different times, it is possible that 
the stabilization mechanism happens after the end of inflation and reduces the separation of 
the two branes. However, if $z_c$  at the onset of inflation does not 
satisfy: $ e^{\gamma N} \ll \alpha$, inflation would eventually destabilize 
the configuration of the two branes and the perturbative approach would break down.  
We also make no comments on how the inflation is stopped, however, we believe that the
end of slow-roll conditions eventually leads to graceful exit of inflation. Though it would be 
interesting to investigate the possibility of the stabilization mechanism as 
a way to stop inflation and generate reheating, it is beyond the scope of this work.   
Similar expression can be derived for $\Phi^{+}(\tau_f)$ in terms of the number of e-foldings $N$,
\begin{eqnarray}
\label{pslov8}
\frac{\Phi^{+}(\tau_f)}{G_{P}}=1-\frac{e^{-\gamma N}}{\alpha}\,.
\end{eqnarray}
In the perturbative limit, $f$, the radion field determines the perturbation of the separation between 
the two branes. The result in  Eq.(\ref{psolv7}) shows, that by assuming
$f=0$ 
as an initial condition for inflation to occur, $f(\tau)$ increases during inflation. 
The increased value depends on the number
of e-foldings that can be achieved during inflation. Therefore, inflation on the Planck 
brane results in increasing the separation between the two branes, i.e, the size of 
the AdS space. 
On the other hand
Eq.(\ref{pslov8}) suggests that the increase in the radion field $f(\tau)$
also
leads to gradual increase in $\Phi^{+}$ and at the end of 
inflation it's value asymptotically approaches  $G_{\rm P} \sim 1/(kG_5)$.    

\subsection{On the negative tension brane} 

In this section we study the feasibility of an inflationary solution on 
the negative tension brane. The action on the negative tension brane 
is given in Eq.(\ref{S}) and Eq.(\ref{Smatter}) with negative superscript. 
Again in the perturbative limit, $\Phi^{-}$ in Eq.(\ref{Phi}), reduces to a simple form
\begin{eqnarray}
\Phi^{-} \approx \frac{1}{kG_5} \alpha e^{2k(\alpha-1)f}\,,
\end{eqnarray}
and the parameter:
\begin{eqnarray}
\label{parm}
\omega_{-} \approx -\frac{3}{2}(1+ \frac{2}{\alpha}-\frac{3}{kG_5\Phi^{-}})\,.
\end{eqnarray}
Within the perturbative limit $e^{2k\alpha f} \ll \alpha $, $2\omega_{-}+3$ is greater than zero, 
provided
\begin{eqnarray}
\label{co}
f(\tau) < \frac{\ln(3/2)}{2k\alpha }\,.
\end{eqnarray}
In the perturbative limit, we also have:
\begin{eqnarray}
\label{negative}
kG_5 \Phi^{-} \approx \alpha \gg 1\,,
\end{eqnarray}
for $f \sim 0$. 
Note, that the situation is exactly opposite to that of the positive tension brane, where $\Phi^{+}$ is 
small and $\omega_{+}\gg 1$.  

With these assumptions and with the help of slow-roll conditions Eqs.(\ref{slowroll1}--\ref{slowroll3}), 
Eq.(\ref{eom1}) can be simplified to yield, 
\begin{equation}
\label{nsolv1}
3H^2 \approx \frac{8\pi V_0}{\Phi^{-}}\,. 
\end{equation}
Here we have used Eq.(\ref{parm}), and we have also assumed the initial
condition $f(\tau=\tau_0) \sim 0$. 
On the other hand Eq.(\ref{eom2}), around $f\sim 0$, reduces to:
\begin{equation}
\label{nsolv2}
\frac{\dot{\Phi}^{-}}{\Phi^{-}} \approx \frac{16\pi}{27}
\frac{kG_5 V_0}{H(1-G_{\phi}\Phi^{-})}, 
\end{equation}
where $G_{\phi}\equiv \frac{2}{3}\frac{kG_5}{\alpha}$, such that around 
$f\sim 0$, $G_{\phi}\Phi^{-} \sim 2/3$.
Hence, the slow-roll condition: $\dot{\Phi}^{-}/\Phi^{-} \ll H$ requires  
$kG_5 V_0/H \ll H$. However, it is easy to show that combined with the result of 
Eq.(\ref{nsolv1}) this leads to 
an inconsistent result $kG_5 \Phi^{-} \ll 1$, compared to  
Eq.(\ref{negative}). 
It leads to the conclusion that within the perturbative limit the
slow-roll inflation on the negative tension brane breaks 
down from the very beginning.

The result can be understood as follows. On the negative tension brane, 
the effective physical scale is reduced by a factor of $\alpha \equiv e^{2kz_c}$ 
compared to the fundamental scale $M_{Pl}$, the same mechanism used by Randall-Sundrum to 
explain the hierarchy between the Planck scale and the electroweak scale \cite{rs}.
 Hence, inflation is driven by the flat potential $V_0/\alpha$ of $\chi$ Eq.(\ref{nsolv1}).
On the other hand, due to the fact that $\omega_{-}$ is small on the negative brane
, in fact $2\omega_{-}+3\sim 1/\alpha$, the coupling between $\Phi^{-}$ and the inflaton $\chi$  
is boosted  by a factor of $\alpha$ (see the term on the right-hand of Eq.(\ref{eom2})). 
This creates an effective potential for $\Phi^{-}$ with a steep slope, 
hence the evolution of $\Phi^{-}$ is  
 no more subdominant compared to the Hubble expansion rate which is set here 
by the reduced strength of $V_0$. And eventually the slow-roll assumptions
break down.  
On the other hand, one may also conclude that the inflation quickly pushes the two branes 
out of their equilibrium positions, hence, destabilizes the configuration.   

Note that the inconsistency is related to the fact that the value of $\omega_{-}$ 
in the perturbative limit is non-physical, i.e., its value does not satisfy the present 
experimental bound $\omega > 3000$ as we discussed earlier. One may consider 
a situation where the stabilization process is inter-winded with the inflation. 
By including new contributions from massive particles in the bulk, the effective 
steep potential of $\Phi^{-}$ could be balanced to provide a possible
solution to the problem. (It was suggested by some of the studies on 
cosmology in Randall-Sundrum scenario \cite{csaki}.) 
Inflation in such a case would be driven by the massive 
Kaluza-Klein modes 
of the bulk field, which could possibly assist inflation \cite{assist} 
on the brane, work in this direction is in progress.

\section{Discussion}

We have shown that it is possible to discuss the dynamics of inflation with additional matter 
Lagrangian on the positive tension brane. Assuming inflation takes place before the stabilization of the 
brane separation, we find inflation on the positive tension brane  
displaces the radion field, but can stay within the perturbative limits with large brane separations. 
Due to the expansion of the positive
tension brane, the separation between the two branes increases slightly and this could 
be the initial condition for a subsequent process in which the brane
separation is stabilized and the radion field becomes massive.  
However, on the negative tension brane, the slow-roll inflation destabilizes the brane separation 
at the very beginning, and it
is not possible to reach the general relativity limit for a given separation between the two branes.

\section*{Acknowledgments}
J.W. would like to thank M. Cveti\v c and D. Chung for helpful 
discussions. 
A.M. is supported by the Inlaks foundation and an ORS award. J.W. is supported by U.S.\ Department 
of Energy Grant No.~DE-AC02-76CH03000.


\begin{references}  
\bibitem{nima} N. Arkani-Hamed, S. Dimopoulos and G. Dvali, Phys. Lett. B. {\bf 246}, 377
(1990); Phys. Rev. D. {\bf 60}, 104002 (1990); I. Antoniadis. S. Dimopoulos and G. Dvali,
Nucl. Phys. B {\bf 516}, 70 (1999); N. Arkani-Hamed, S. Dimopoulos and J. March-Russell, hep-th/9809124/

\bibitem{lykken} L. J. Hall and D. Smith, Phys. Rev. D {\bf 60}, 085008
(1999); T. G. Rizzo, Phys. Rev. D {\bf 59}, 115010 (1999); T. Han, J. D.
Lykken and R.-J. Zhang, Phys. Rev. D {\bf 59}, 105006 (1999); G. F. Giudice, R. Rattazi
and J. D. Wells, Nucl. Phys. B {\bf 544}, 3 (1999); E. A. Mirabelli, M. Perelstein and
M. E. Peskin, Phys. Rev. Lett. {\bf 82} 2236 (1999); J. L. Hewett, Phys. Rev. Lett.
{\bf 82} 4765 (1999).

\bibitem{numero} D. H. Lyth, Phys. Lett. B {\bf 448} 191 (1999); N. Kaloper and A. Linde,
Phys. Rev. D. {\bf 59} 101303 (1999); A. Mazumdar, Phys. Lett. B. {\bf 469}, 55 (1999);
T. Nihei, Phys. Lett. B {\bf 465}, 81 (1999); E. Halyo, Phys. Lett. B {\bf 561}, 109
(1999); 
R. Maartens, D. Wands, B. A. Bassett and I. Heard, hep-ph/9912464; N. Kalpoer,
A. R. Liddle, hep-ph/9910499.

\bibitem{rs} L. Randall and R. Sundrum Phys. Rev. Lett. {\bf 83}, 3370 (1999);
Phys. Rev. Lett. {\bf 83}, 4690 (1999).

\bibitem{charmousis} C. Charmousis, R. Gregory and V. A. Rubokov, hep-th/9912160.

\bibitem{gw} W. Goldberger and M. Wise, hep-ph/9911457; Phys. Rev. Lett. {\bf 83}, 4922 (1999). 

\bibitem{chiba} T. Chiba, gr-qc/0001029.

\bibitem{adkm:inflation} N. Arkani-Hamed, S. Dimopoulos, N. Kaloper, J. March-Russell, hep-ph/9903239; hep-ph/9903224. 


\bibitem{csaki} G. Dvali, S. H. H. Tye, Phys. Lett. B {\bf 450}, 72
(1999);
C. Csaki, M. Graesser and J. Terning, Phys. Lett. B {\bf 456} (1999);
C. Csaki, M. Graesser, L. randall, J. Terning, hep-ph/9911406; J. M.
Cline,
Phys. Rev. D {\bf 61} 023513 (2000); E. E. Flanagan, S. H. H. Tye and I.
Wasser, hep-ph/9909373; A. Riotto. hep-ph/9904485;  H. B. Kim and H. D. Kim
, Phys.Rev. D{\bf 61},064003 (2000). 

\bibitem{TS}T. Shiromizu, gr-qc/9910076. 
     

\bibitem{barrow0} C. Brans and R. H. Dicke, Phys. Rev. {\bf 124}, 925 (1961);
  P. G. Bergmann, Int. J. Theor. Phys. {\bf 1}, 25 (1968); R. V. Wagoner, Phys. Rev.
  D {\bf 1}, 3209 (1970); C. M. Will, {\it Theory and experiment in gravitational physics},
   Cambridge University Press, Cambridge, (1993).

\bibitem{nordt} K. Nordvedt, Astrophys. J. {\bf 161}, 1059 (1970).

\bibitem{will} C. M. Will, gr-qc/9811036.

\bibitem{barrow} J. D. Barrow, Phys. Rev. D {\bf 51}, 2729 (1995).

\bibitem{assist} A. R. Liddle, A. Mazumdar and F. Schunck, Phys. Rev. D {\bf 58}, 061301 (1998);
E. J. Copeland, A. Mazumdar and N. J. Nunes, Phys. Rev. D {\bf 60}, 083506 (1999).

\end{references}
\end{document}